\documentclass[a4paper]{jpconf}
\usepackage{graphicx}
\begin{document}
\title{Update on Chiral Symmetry Restoration in the Context of Dilepton Data}

\author{Ralf Rapp}

\address{Cyclotron Institute and Department of Physics and Astronomy, 
Texas A\&M University, College Station, TX 77843-3366, U.S.A.}

\ead{rapp@comp.tamu.edu}

\begin{abstract}
We evaluate currently available information on low-mass dilepton and
direct-photon emission spectra measured in ultrarelativistic heavy-ion 
collisions. In the first part an attempt is made to develop a consistent 
picture of the in-medium effects on the electromagnetic spectral function 
and pinpoint its emission history by utilizing its radial and elliptic flow 
signatures. In the second part we elaborate on the implications of the 
empirical information on the nature of chiral symmetry restoration. We 
indicate how the melting of the $\rho$ resonance in hot and dense matter is 
related to, and compatible with, the reduction of chiral order 
parameters as ``measured" in thermal lattice QCD. 
\end{abstract}

%%%%%%%%%%%%%%%%%%%%%%%%%
\section{Introduction}
%%%%%%%%%%%%%%%%%%%%%%%%
The spontaneous breaking of chiral symmetry (SBCS) in the QCD vacuum is believed
to be a consequence of the condensation of scalar quark-antiquark pairs.
It manifests itself in the generation of mass in strong interactions and, more 
precisely, in lifting the degeneracies within chiral multiplets ($\pi$-$\sigma$, 
$\rho$-$a_1$, $N$-$N^*(1535)$, ...). The search for (partial) restoration of SBCS 
in hot and dense matter as a fundamental phenomenon is one of the core missions 
of the ultrarelativistic heavy-ion programs at laboratories around the world. 
The most promising observable in this regard are invariant-mass spectra of 
dileptons ($e^+e^-$ or $\mu^+\mu^-$) which open a direct window on the 
in-medium spectral function of the electromagnetic (EM) current, 
cf.~Refs.~\cite{Rapp:2009yu,Tserruya:2009zt,Specht:2010xu,Rapp:2011is,Gale:2012xq} 
for recent reviews. In the vacuum, and at low mass ($M\le 1$\,GeV), the  EM 
spectral function reflects the mass distribution of the light vector mesons $\rho$, 
$\omega$ and $\phi$, and thus the dynamical generation of mass in QCD. Thermal 
radiation of low-mass dileptons is ideally suited to illuminate the changes in 
the vector-meson mass distributions as the (pseudo-) critical temperature for chiral 
restoration, $T_{\rm pc}^\chi\simeq160$~MeV, is approached and surpassed. However, 
robust conclusions from measurements of dilepton spectra in heavy-ion collisions 
require a number of rather challenging steps. First, one needs sufficiently accurate 
measurements of so-called ``excess" radiation (beyond final-state decays) to 
extract its spectral shape. Second, such measurements need to cover a large
range of collision energies to establish systematic trends, enabling the 
extraction of genuine features in the radiation. In particular, one needs to 
correlate the observed spectral shapes with the thermodynamic properties of
the emission sources, most notably its temperature(s). Third, model calculations
and predictions need to be tested against the data which is critical for deducing 
mechanisms underlying observed spectral modifications. Such comparisons not only 
require calculations of the in-medium EM spectral functions but also a good 
control over space-time evolution of the fireball at each collision energy.       
Finally, the model calculations have to be rooted in the bigger picture of
QCD thermodynamics, and specifically in the context of chiral restoration
by utilizing relations between the EM spectral functions and chiral order 
parameters, where the latter can be tested (or extracted) from thermal lattice 
QCD. It is the purpose of this proceedings to give an update on carrying out
these steps. 

The remainder of this article is organized into two main sections. The first 
one (Sec.~\ref{sec_exp}) contains a phenomenological assessment of the current 
experimental situation regarding the extraction of the in-medium EM 
spectral function and the pertinent regimes in temperature and baryon density 
that it corresponds to. The second one (Sec.~\ref{sec_theo}) summarizes how the 
phenomenological findings relate to chiral symmetry restoration, 
%and possibly deconfinement, 
invoking both rigorous and more heuristic arguments. Both sections are not 
divided up any further to stipulate the comprehensive nature of the discussion. 
We briefly conclude in Sec.~\ref{sec_concl}. 

%%%%%%%%%%%%%%%%%%%%%%%%%%%%%%%%%%%%%%%%%%%%%%%%%%%%%%%%%%%%%
\section{Electromagnetic Emission Spectra from Experiment}
\label{sec_exp}
\begin{figure}[!t]
\begin{minipage}{18pc}
\includegraphics[width=18pc]{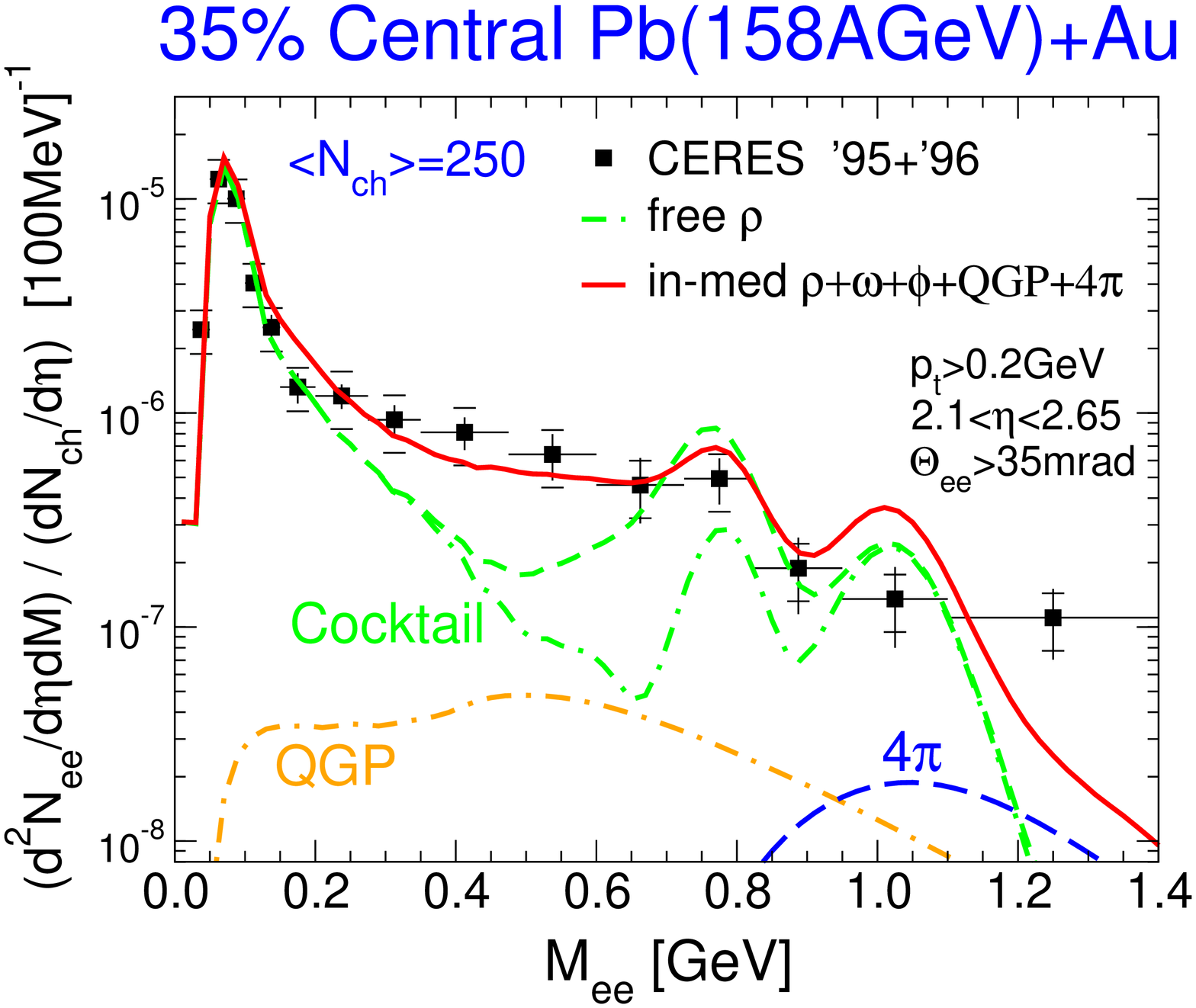}
\end{minipage}\hspace{1pc}%
\begin{minipage}{19pc}
\vspace{0.6pc}
\includegraphics[width=19pc]{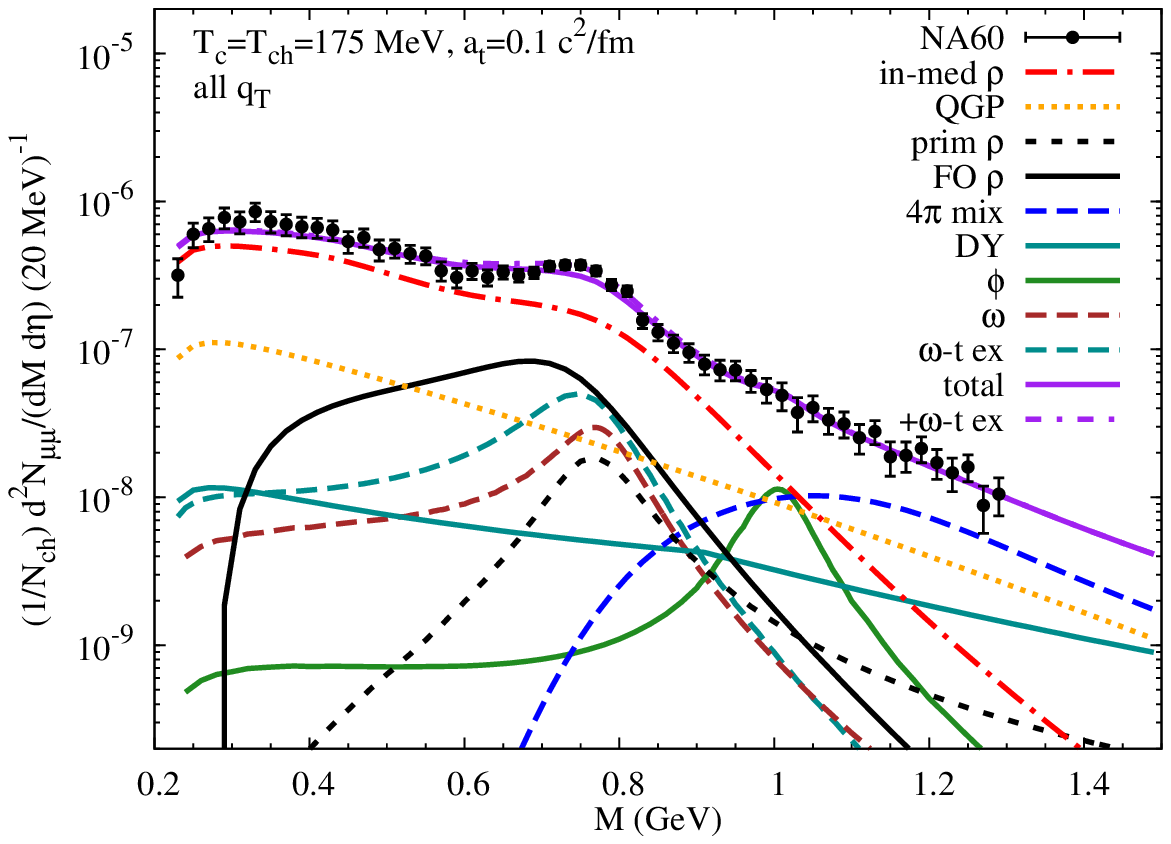}
\end{minipage}
\caption{Dilepton spectra measured at SPS energies ($\sqrt{s}$=17.3\,AGeV) by the 
CERES/NA45 collaboration in Pb-Au collisions (left panel)~\cite{Agakichiev:2005ai}, 
and by the NA60 collaboration in In-In (right panel)~\cite{Arnaldi:2008fw}, compared
to the same theoretical approach~\cite{vanHees:2007th}.
The CERES data are taken in the $e^+e^-$ channel and include acceptance cuts in the 
(di-) electron tracks, while the NA60 data are for $\mu^+\mu^-$, subtracted of the 
cocktail and fully corrected for (di-) muon acceptance cuts.}
\label{fig_sps}
\end{figure}

%%%%%%%%%%%%%%%%%%%%%%%%%%%%%%%%%%%%%%%%%%%%%%%%%%%%%%%%%%%%%
The first unambiguous detection of a low-mass dilepton excess in ultrarelativistic 
heavy-ion collisions (URHICs) was achieved by the CERES/NA45 collaboration, first 
in the S-Au system~\cite{Agakishiev:1995xb} and later with improved precision
in Pb-Au collisions~\cite{Agakichiev:2005ai,Adamova:2006nu}. Early theoretical 
analyses successfully described these data utilizing the conjecture of a dropping 
$\rho$-meson mass in hot and dense matter~\cite{Li:1995qm}, as a direct consequence 
of the reduction of the quark condensate (later, this connection was found to be 
problematic and revisited~\cite{Brown:2009az}). Shortly thereafter, several groups 
started to evaluate ``more mundane" medium modifications, by performing 
``standard" hadronic many-body (or thermal-field theory) calculations of the $\rho$ 
spectral function using {\it known} (or at least well-constrained) hadronic 
interactions in vacuum as an input. These calculations, performed for cold nuclear 
matter~\cite{Chanfray:1993ue,Herrmann:1993za,Friman:1997tc,Klingl:1997kf,Peters:1997va,Urban:1998eg,Cabrera:2000dx},
a hot meson gas~\cite{Haglin:1994xu,Pisarski:1995xu,Song:1996dg,Rapp:1999qu,Ayala:2003yp}, 
or hot and dense hadronic
matter~\cite{Rapp:1995zy,Rapp:1999us,Eletsky:2001bb,Ghosh:2011gs}, generically produce
a rather strong broadening of the in-medium $\rho$ spectral function with little mass
shift, especially in nuclear media. What was originally intended to provide a baseline 
for disentangling more ``exotic" medium effects, subsequently turned into 
a fair description of the CERES dilepton enhancement all by itself, see the left panel
of Fig.~\ref{fig_sps}. Does this imply that dilepton data are not sensitive (or even
unrelated) to chiral restoration? Our answer is no, as will be disucssed in 
Sec.~\ref{sec_theo} below.

\begin{figure}[!t]
\begin{minipage}{19pc}
\includegraphics[width=15.5pc,height=22pc,angle=-90]{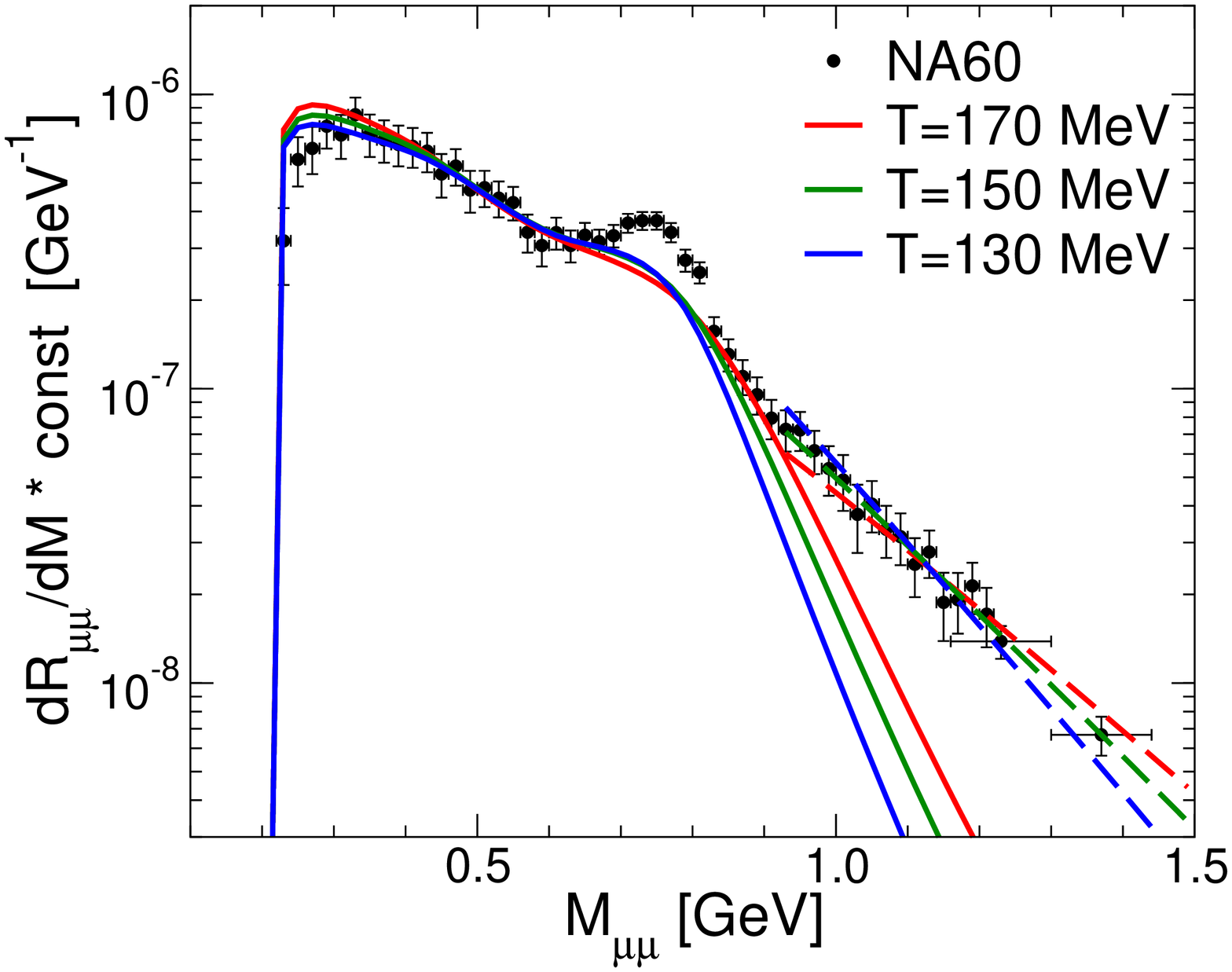}
\end{minipage}
\hspace{0.5pc}%
\begin{minipage}{19pc}
\vspace{0.3pc}
\includegraphics[width=18pc]{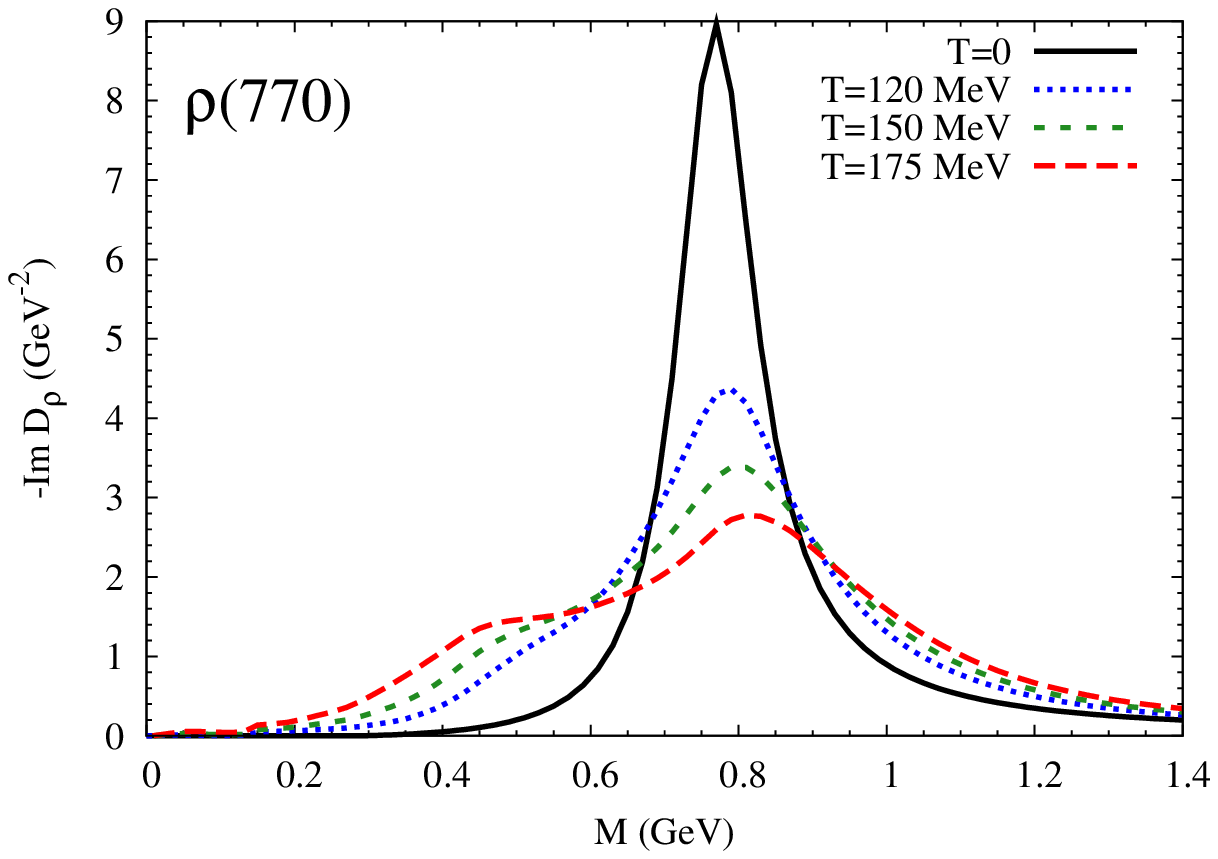}
\end{minipage}
\caption{Left panel: Comparison of the thermal dilepton emission rates at fixed
temperature (as figuring into the full calculation of the spectra shown in 
Fig.~\ref{fig_sps}) with the acceptance
corrected NA60 spectra. The rates are a combination of the low-mass $\rho$ 
contribution (solid lines) and a continuum ``4$\pi$" part (dashed lines) which 
for simplicity has been approximated by a perturbative continuum limited to 
$M$$>$0.9\,GeV.  Both contributions carry the relative normalization as used 
in the fireball evolution (including pion fugacity factors), but are scaled 
by an overall constant to match the data at $M$$\simeq$0.5\,GeV. Isoscalar
contributions ($\omega$ and $\phi$) to the rates are not included.   
Right panel: In-medium $\rho$ spectral function
based on Ref.~\cite{Rapp:1999us} as figuring into the calculations of the dilepton 
spectra in Fig.~\ref{fig_sps}~\cite{vanHees:2007th} and underlying the low-mass
in-medium hadronic rates in the left panel.}
\label{fig_rate}
\end{figure}

A quantitative test of the ``$\rho$-broadening" became possible with the NA60 dimuon 
spectra in In-In($\sqrt{s}$=17.3\,AGeV) collisions~\cite{Arnaldi:2006jq,Arnaldi:2008fw}, 
which have set a new standard for precision in dilepton spectroscopy in URHICs to date, 
see right panel of Fig.~\ref{fig_sps}. These data have been fully corrected for the 
detector acceptance, which, for the first time in URHIC dilepton spectroscopy, renders 
the mass spectra (Lorentz-) {\em invariant}. This means that their shape is 
{\it unaffected} by the blue shift due to the collective flow of the expanding fireball, 
and thus they directly reflect the spectral distribution of the medium's emission rate 
(in addition, 
the excellent mass resolution and statistics allowed for a subtraction of the hadronic 
decay cocktail, defining the notion of ``excess" spectra). If the emission emanates from 
a locally thermalized medium, it takes the well-known form
\begin{equation}
\frac{dN_{ll}}{d^4xd^4q} = -\frac{\alpha_{\rm em}^2}{\pi^3 M^2} \
       f^B(q_0;T)~{\rm Im}\Pi_{\rm em}(M,q;\mu_B,T) \ , 
\label{rate}
\end{equation}
where ${\rm Im}\Pi_{\rm em}$ is the in-medium EM spectral function, $f^B$ the thermal
Bose distribution and the factor $1/M^2$ is a remnant of the intermediate 
propagator of the virtual photon. 
The predictions of hadronic many-body theory (evolved over a thermal fireball evolution
constrained by hadron spectra)~\cite{vanHees:2006ng,vanHees:2007th} turn out to agree 
well with these data, cf.~right panel of Fig.~\ref{fig_sps} (see also 
Refs.~\cite{Dusling:2006yv,Ruppert:2007cr,Santini:2011zw,Linnyk:2011hz}).
This actually provides non-trivial information beyond the realm of strict reliability
of hadronic theory, which we estimate to be up to $T\simeq150$\,MeV, where the total
hadron density reaches about 2$\varrho_0$ ($\varrho_0$=0.16\,fm$^{-3}$ denotes
nuclear matter saturation density). A good portion of the low-mass dilepton enhancement 
is radiated from temperatures around and somewhat below $T_{\rm pc}$ in the fireball 
evolution. In other words, the NA60 data quantitatively support a smooth extrapolation 
of the hadronic $\rho$ broadening into the temperature region of the chiral transition.
%This indeed implies that the $\rho$ width  a melting of the resonance. 
More explicitly, the observed spectra follow from a convolution of the above
rate, Eq.~(\ref{rate}), over 3-momentum and 4-volume of the expanding medium, 
\begin{eqnarray}
\frac{dN_{ll}}{dMdy}
&=& \frac{1}{\Delta y} \int\limits_{\tau_0}^{\tau_{\mathrm{fo}}}
d\tau \int\limits_{V_{\mathrm{FB}}} d^3x \int \frac{Md^3 q}{q_0} \
\frac{dN_{ll}}{d^4 xd^4q}(M,q;T(\tau),\mu_B(\tau)) 
\label{int}
\\
&\simeq& \frac{V_4}{\Delta y} \int \frac{d^3q}{Mq_0} 
\frac{\alpha_{\rm em}^2}{\pi^3}f^B(q_0;\bar{T}) \  (-{\rm Im}\Pi_{\rm em}(M,q;\bar{T}, \bar{\mu}_B)) \ .
\label{spec}
\end{eqnarray}
In the second line we have condensed the space-time integration into a 4-volume, $V_4$,
at the expense of replacing the time-dependent temperature and baryo-chemical potential
(or baryon density) in the dilepton rate by average values, $\bar{T}$ and $\bar{\mu}_B$, 
respectively. The latter can be extracted from the model calculations which describe the 
data to be $\bar{T}\simeq$~150-160\,MeV and $\bar{\mu}_B^{\rm tot}\simeq$~250-300\,MeV 
(or $\bar{\varrho}_B^{\rm tot} \simeq$~0.7-1\,$\varrho_0$). The remaining 3-momentum 
integral in Eq.~(\ref{spec}) is a Lorentz scalar and can thus be evaluated in the local 
rest frame. This is routinely done in model calculations of the spectral function, as
shown in the left panel of Fig.~\ref{fig_rate}, including the dimuon phase-space factor 
and mass threshold, $M_{\rm thr}=2m_\mu=211$\,MeV. One immediately recognizes a 
remarkable reminiscence of the theoretical rates, calculated a decade 
ago~\cite{Rapp:1999us}, with the NA60 data. 
This corroborates the possibility of extracting an ``average" $\rho$-meson width 
which turns out to be $\bar\Gamma_\rho^{\rm med}\simeq$~350-400\,MeV (which is not
not far from the value found for cold nuclear matter at saturation density). This
necessarily implies larger widths in the earlier stages of the fireball evolution
(later and earlier contributions are, in fact, required to properly account for
the total excess yield), which reach around $\Gamma_\rho(T_{\rm pc})\simeq$~600\,MeV, 
before a transition to
QGP rates is performed. At this point, the QGP and hadronic rates are very similar
(see left panel of Fig.~\ref{fig_rate}), which has been interpreted as quark-hadron
duality in dilepton rates across $T_{\rm pc}$~\cite{Rapp:1999if}. A similar feature
was found for photon rates in Ref.~\cite{Kapusta:1991qp}, and later again in
Ref.~\cite{Turbide:2003si}. 

An important question is whether one can obtain independent information on the
medium's temperature from which the observed spectral shape is originating. As 
stated above, invariant-mass spectra are an ideal tool to evaluate emission
temperatures due to the absence of blue shifts induced by the collective flow (which 
are present in transverse-momentum ($q_T$) spectra). However, as evident from 
Eq.~(\ref{spec}), the thermal slope associated with the Bose distribution is modulated 
by the medium effects in the EM spectral function. Nevertheless, when using the 
in-medium hadronic rates underlying the description of the NA60 data in Fig.~\ref{fig_sps} 
(right) as a ``thermometer", one finds that the slope in the data over the 
range $M=$~0.3-1.5\,GeV is reasonably well reproduced with $T\simeq$~150-170\,MeV, 
cf.~left panel of Fig.~\ref{fig_rate}. Closer inspection of this comparison reveals 
that the enhancement at the very low-mass end, close to the dimuon threshold,
is slightly overestimated, suggesting that in this regime emission from later
stages prevails, where the medium effects are smaller, cf.~the in-medium $\rho$
spectral function in the right panel of Fig.~\ref{fig_rate}. Alternatively, carrying 
out the slope analysis for masses above $M=$~1\,GeV, where the medium 
effects on the rate are less pronounced (resembling a structureless continuum), one 
deduces a slope of around $T\simeq$~170\,MeV up to $M\simeq 1.5$\,GeV, with a tendency 
to further increase at still higher mass. This is consistent with the well-known 
feature that at higher mass (or higher $q_T$, as governed by the energy, $q_0$,  
figuring into the thermal distribution function), the temperature sensitivity of 
the thermal exponential increasingly biases the contributions to the spectra toward 
earlier phases (i.e., higher $T$)~\cite{Rapp:2011is}. The NA60 collaboration has 
also conducted a systematic investigation of slope parameters, $T_{\rm eff}$, of 
$q_T$ spectra in various mass bins~\cite{Arnaldi:2007ru}. Here, $T_{\rm eff}$ 
contains the radial-flow blue shift,
schematically as $T_{\rm eff} \simeq T + M \beta_{\rm av}^2$, where 
$\beta_{\rm av}$ denotes the average expansion velocity of the fireball
at a given moment. The extracted values for $T_{\rm eff}$ gradually increase
from about 170\,MeV at threshold up to ca.~260\,MeV at the $\rho$ mass, 
decreasing thereafter and leveling off at ca. 200-220\,MeV for $M>1$\,GeV. 
This is remarkably consistent with the information from $M$-spectra, i.e., an 
emission source of hot hadronic matter at and below the $\rho$-mass with a 
mass-dependent slope increase, and emission from around $T_c$ and higher (with a 
possibly large QGP component) above the $\rho$ mass.     
 
An intriguing aspect of the $\rho$-broadening as obtained from many-body theory 
in hot and dense hadronic matter is the prevalence of baryon-induced medium 
effects~\cite{Rapp:1995zy,Rapp:1999us}. Theoretically, this is a consequence of
quantitative constraints on the $\rho$-meson coupling to nucleons as deduced, 
e.g., from nuclear photoabsorption data on the proton and on 
nuclei~\cite{Rapp:1997ei}. This feature raised some doubts in the early stages 
of interpreting the CERES data, as the experimental pion-to-baryon ratio at 
full SPS energy is about 5:1. However, many of the pions observed in the 
final state originate from resonance decays; including these in a thermally 
equilibrated ``hadron resonance gas" gives meson-to-baryon ratios of closer to 
2:1. Nonetheless, the question arises how the properties of the excess radiation 
develop with the nuclear collision energy, i.e., with the relative baryon content 
in the system. The first step in this direction was a CERES measurement at a lower 
SPS bombarding energy of $E_{\rm lab}$=40\,AGeV ($\sqrt{s}$=8.7\,AGeV)~\cite{Adamova:2003kf}. 
While the pion multiplicity decreases by ca.~40\%, the low-mass dilepton enhancement 
over the cocktail indicates an increase over the result at 158\,AGeV, consistent 
with the anticipated prevalence of baryon-driven medium effects. The predictions 
from the in-medium broadened $\rho$ spectral function describe the data well. This 
trend continues all the way down to relativistic energies in the BEVALAC/SIS regime 
($E_{\rm lab}$=1-2\,AGeV), where a large excess radiation has also been 
reported~\cite{Porter:1997rc,Agakishiev:2011vf}. The next step is to go to higher 
energies as available at the colliders RHIC and LHC. Here, the net baryon density at 
mid-rapidity becomes small. It was pointed out, however, that at temperatures close 
to $T_{\rm pc}$, the {\em sum} of the densities of baryons ($B$) and anti-baryons 
($\bar{B}$) is appreciable, about 0.7$\varrho_0$, and thus critical in producing a 
large broadening of the $\rho$, since it interacts equally with baryons and 
anti-baryons due to $CP$ invariance~\cite{Rapp:2000pe}. Since the concept of chemical 
freeze-out continues to hold in nuclear collisions at RHIC energy, the total number 
of baryons plus anti-baryons does not change appreciably until thermal freeze-out at
$T_{\rm fo}$$\simeq$100\,MeV, and thus their density drops much slower than would be 
the case in chemical equilibrium (where it would decrease dramatically due to 
the large mass penalty on baryons). The upshot of this discussion is that the hadronic 
in-medium effects at collider energies were predicted to be comparable to that at SPS 
energies, and thus the low-mass dilepton enhancement at RHIC and LHC is expected to be
quite similar in magnitude and shape to what has been observed at SPS. The PHENIX
data of Ref.~\cite{Adare:2009qk} do not support this expectation: a large enhancement
in central Au-Au($\sqrt{s}$=200\,AGeV) has been reported which cannot be described
by the in-medium hadronic effects as is the case at the SPS. One should note, however,
that the enhancement in non-central collisions is significantly smaller. A new 
mechanism should thus be operative in central Au-Au at RHIC, which does not 
prominently figure at SPS, nor in more peripheral collisions at RHIC. On the other 
hand, the STAR collaboration has reported preliminary data for dielectrons in central 
Au-Au($\sqrt{s}$=200\,AGeV)~\cite{Zhao:2011wa}, which indicate a much smaller 
enhancement, not incompatible with the effects expected from the in-medium $\rho$ 
broadening. At the recent Quark Matter 2012 meeting, the STAR collaboration went
another step further, presenting a systematic dielectron measurement from the
RHIC beam-energy scan program~\cite{Geurts:2012}. A persistent low-mass enhancement 
was found in Au-Au at collisions energies of $\sqrt{s}$=19.6, 39, 62 and 200\,AGeV. 
The invariant-mass spectra at the lowest energy (19.6\,GeV) exhibited the largest
excess over the cocktail, and agree very well with the CERES data in 
Pb-Au($\sqrt{s}$=17.3\,AGeV). The predictions from hadronic many-body theory 
(with a moderate QGP portion in the low-mass regime) show good agreement with 
this excitation function. The STAR measurements (together with the CERES and
NA60 data) thus suggest a universal origin of the low-mass dilepton enhancement 
in URHICs from $\sqrt{s}\simeq$~10-200\,GeV. New
data reported by the PHENIX collaboration for peripheral and semi-central
Au-Au, taking advantage of the hadron blind detector (HBD), also support this
scenario~\cite{Tserruya:2012}. 

The STAR collaboration furthermore presented first measurements
of the dielectron elliptic flow, $v_2$. Within the currently rather large 
uncertainties of this very challenging measurement, the $v_2$ of the 
dielectrons in the low-mass region, divided up into several mass bins, 
was found to be compatible with the simulated $v_2$ of the decay cocktail (mostly
due to the long-lived $\pi$, $\eta$, $\omega$ and $\phi$ contributions). At face
value, this result implies that the excess radiation carries a $v_2$ which is
as large as the hadrons decaying after thermal freezeout. This assertion is, in 
fact, very consistent with the PHENIX measurement of direct photon $v_2$ in Au-Au 
collisions~\cite{Adare:2011zr}, which was found to be compatible with those of 
pions for $q_T\le$3\,GeV. Such an observation is difficult to account for through 
radiation which is dominated by early QGP 
emission~\cite{Liu:2009kta,Holopainen:2011pd,Dion:2011pp}.
This discrepancy can be noticably reduced in a rather straightforward fireball 
scenario where most of the 
bulk $v_2$ is built up by the time the system reaches the phase transition region, 
in connection with photon emission rates for QGP and hadronic matter which continuously 
evolve across $T_{\rm pc}$~\cite{vanHees:2011vb}. The latter feature is indeed satisfied 
when merging hadronic many-body calculations for photon production~\cite{Turbide:2003si} 
with QGP rates in a complete leading-order perturbative evaluation~\cite{Arnold:2001ms}.       
At the same time, a fireball evolution with fully built-up elliptic flow and fairly 
large radial flow at $\sim$$T_{\rm pc}$, is supported empirically by the systematics of 
the measured spectra and $v_2$ of multistrange particles ($\Omega^-$, $\phi$ and $\Xi$), 
as well as by the constituent quark-number scaling of light and strange hadrons; it 
is possible to realize these features in explicit hydrodynamic 
simulations~\cite{He:2011zx}. 
Furthermore, the PHENIX collaboration extracted an inverse slope of their excess 
direct-photon spectra (defined by subtracting the primordial contribution from hard 
$NN$ collisions; late decays, such as $\pi^0,\eta\to\gamma\gamma$, are already taken 
out in the definition of ``direct" photons). It was found to be 
$T_{\rm eff}\simeq(221\pm19^{\rm stat}\pm19^{\rm sys})$\,MeV. This is a rather soft 
slope given its approximate decomposition into a true medium temperature and the 
blue-shift effect on massless particles due to an average radial flow velocity, 
$T_{\rm eff}\simeq T\sqrt{(1+\beta_{\rm av})/(1-\beta_{\rm av})}$. For example, 
for an average flow velocity of $\beta_{\rm av}=0.3(0.4)$, the blue-shift 
correction amounts to $\sim$35(50)\%. This further corroborates that the 
prevalent emission of the photons measured by PHENIX should be around 
$T_{\rm pc}$.
%\footnote{Alternative explanations of these findings have been suggested, 
%e.g., in Refs.~\cite{Basar:2012bp} associated with primordial magnetic fields. However, it 
%remains a challenge to simultaneously and quantitatively describe  the ``large" $v_2$,  
%``small" slope and ``large" yield from an early source.}

To summarize this section, the excess of EM radiation observed in URHICs to date is
remarkably consistent with a thermal source of dileptons and photons from a 
hydrodynamically evolving medium, and thus naturally fits into the current ``standard
model" of these reactions. Employing state-of-the-art emission rates allows for an 
accurate description of precision dilepton data at the SPS, and accounts for
data available at both lower SPS energy and the most the recent spectra obtained
in a first systematic energy scan in the collider regime of RHIC. Slope analyses of
both mass and $q_T$ spectra, as well as their large $v_2$, give strong indications
for this excess radiation to emanate from around the phase transition 
temperature predicted by thermal lattice QCD. The similarity in magnitude and
spectral shape of the excess radiation over the now available formidable range in
collision energy suggests a universal origin of the observations (and further
points to emission around $T_{\rm pc}$). In the following section we will revisit
how microscopic mechanismis of in-medium $\rho$ broadening, which is consistent 
with the data, relates to chiral symmetry restoration in the medium.

%\subsection{Dilepton Mass Spectra: In-Medium Spectral Function}
%\subsection{Transverse-Momentum Spectra and Elliptic Flow: Emission Profile}
%\subsection{Emission off Ground-State Nuclei}

%%%%%%%%%%%%%%%%%%%%%%%%%%%%%%%%%%%%%%%%%%%%%
\section{Implications for Chiral Restoration}
\label{sec_theo}
%%%%%%%%%%%%%%%%%%%%%%%%%%%%%%%%%%%%%%%%%%%%%
\begin{figure}[!t]
\begin{minipage}{19pc}
\includegraphics[width=19pc]{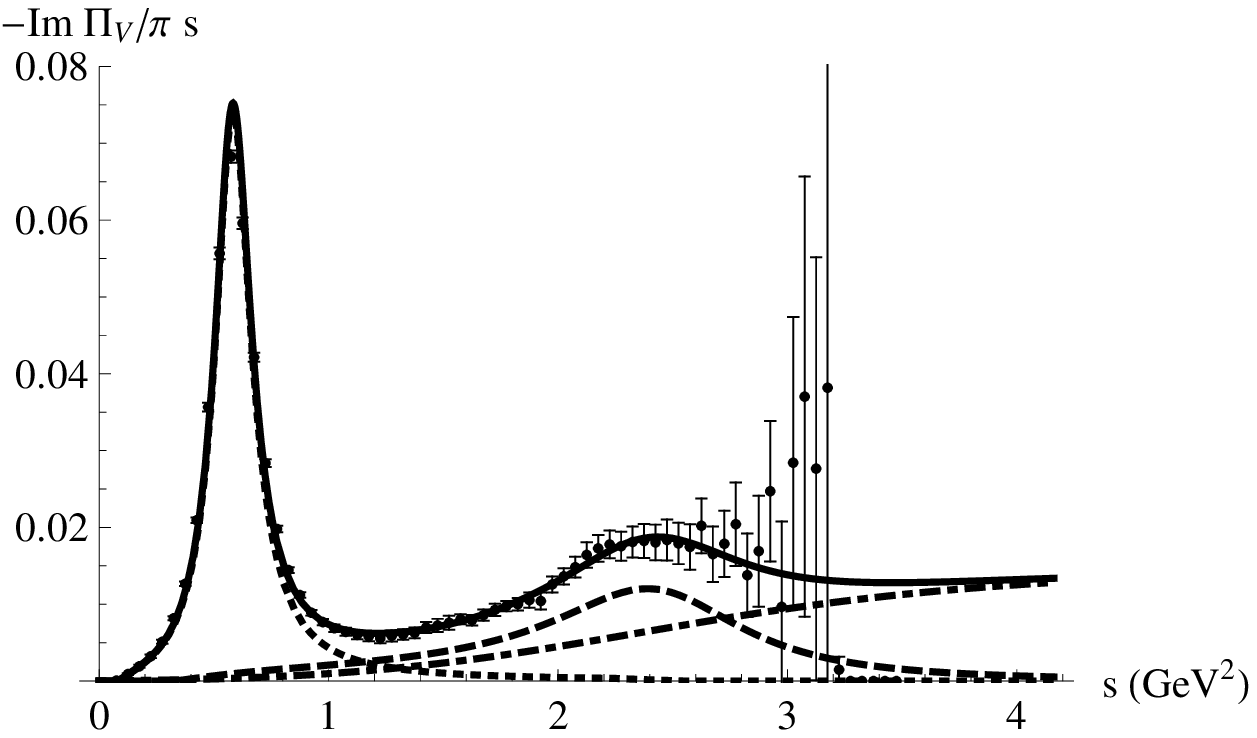}
\end{minipage}\hspace{1pc}%
\begin{minipage}{19pc}
\includegraphics[width=19pc]{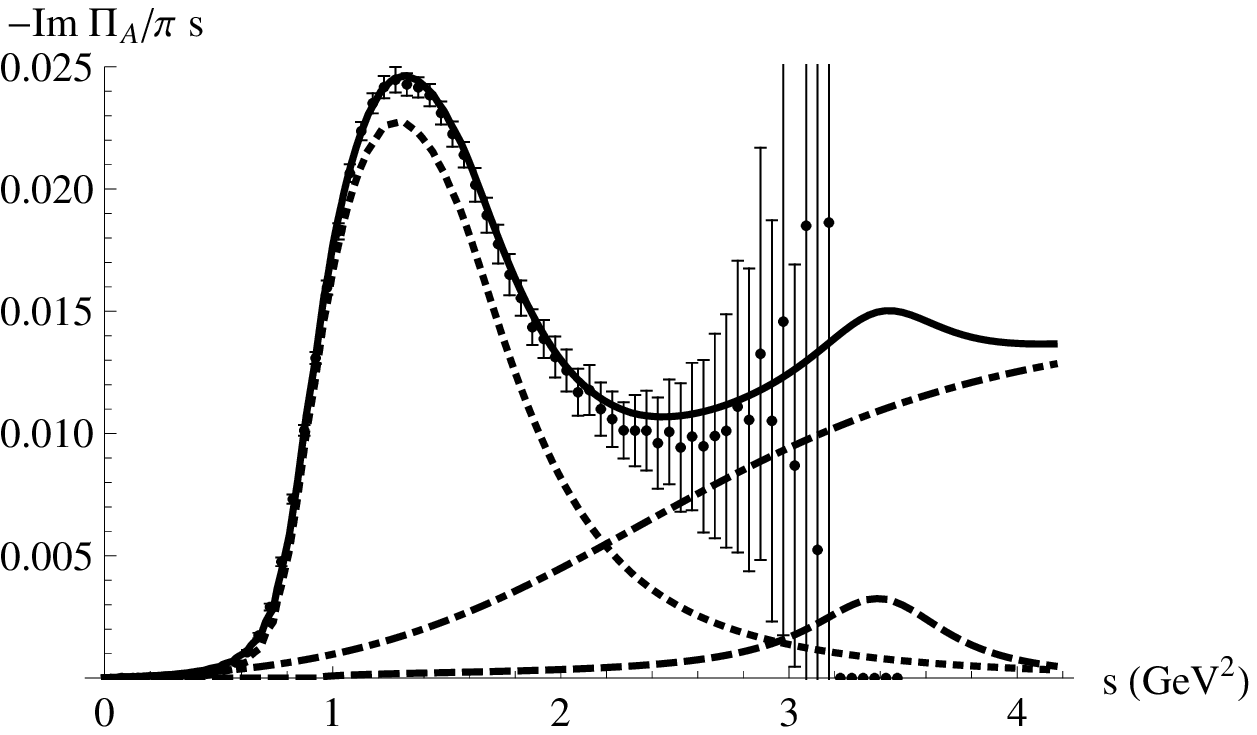}
\end{minipage}
\caption{Vector (left panel) and axialvector (right panel) spectral functions
in vacuum~\cite{Hohler:2012xd}. Notable features are the inclusion of excited 
states $\rho'$ and $a_1'$ (long-dashed lines) and a universal perturbative 
continuum (dash-dotted lines) which starts at significantly higher energies 
than in most previous sum-rule analyses. 
}
\label{fig_vac}
\end{figure}

Rigorous connections between chiral order parameters and the $\rho$ spectral function
(or more precisely: vector-isovector spectral function) can be made via well-known 
sum rule techniques. These are usually divided into two classes, namely the QCD sum 
rules (QCDSRs)~\cite{Shifman:1978bx} and Weinberg (or chiral) sum rules 
(WSRs)~\cite{Weinberg:1967,Das:1967ek}. 
The former are formulated in a given hadronic channel and utilize a 
dispersion integral to relate the physical spectral function to an expansion in 
spacelike momentum transfer with coefficients governed by quark and gluon 
condensates (operator-product expansion). For the vector channel, one has 
\begin{equation}
\frac{1}{M^2}\!\int_0^\infty \!ds \frac{\rho_V(s)}{s} e^{-s/M^2}
= \frac{1}{8\pi^2} \left(1+\frac{\alpha_s}{\pi}\right)
+\frac{m_q \langle\bar{q}q\rangle}{M^4}
+\frac{1}{24 M^4}\langle\frac{\alpha_s}{\pi} G_{\mu\nu}^2\rangle
- \frac{56 \pi \alpha_s}{81 M^6}  \langle \mathcal{O}_4^V \rangle \ldots \, ,
\label{qcdsr}
\end{equation} 
where $\rho_V=-{\rm Im \Pi_V} / \pi$ is the spectral function, $\langle\bar{q}q\rangle$
the chiral condensate, $\langle\frac{\alpha_s}{\pi} G_{\mu\nu}^2\rangle$ the gluon
condensate and $\langle \mathcal{O}_4^V \rangle$ the vector 4-quark condensate. 
The WSRs are, in a sense, more closely related to chiral symmetry
(breaking). They involve energy-weighted integrals (or moments) over the difference 
of vector and axialvector spectral functions, which directly result in order
parameter of chiral symmetry breaking, such as the quark condensate, pion decay
constant or chirally-breaking part of the 4-quark condensate. They read
\begin{eqnarray}
  f_n =  \int\limits_0^\infty ds \ s^n \
  \left[\rho_V(s) - \rho_A(s) \right]  \ ,
  \qquad \qquad \qquad
\label{wsr}
\\
f_{-2} = f_\pi^2  \frac{\langle r_\pi^2 \rangle}{3} - F_A \ , \quad
f_{-1} = f_\pi^2 \ , \quad
f_0   = f_\pi^2 m_\pi^2 \ ,  \quad
f_1 = -2\pi \alpha_s \langle {\cal O}_4^\chi \rangle  \
\label{fn}
\end{eqnarray}
($r_\pi$: pion charge radius, $F_A$: coupling constant for the radiative
pion decay, $\pi^\pm\to\mu^\pm \nu_\mu \gamma$, $\langle {\cal O}_4^\chi
\rangle$: chirally breaking 4-quark condensate), and are referred
to as WSR-0 through -3.  They have been shown to remain valid in the 
medium~\cite{Kapusta:1993hq}. This is quite a fortunate situation as  
the in-medium vector spectral function is the only one readily available 
from experiment.

A quantitative use of these sum rules in medium requires to first establish 
their accuracy and sensitivity in vacuum. This has recently been revisited 
by simultaneously analyzing both sum-rule types in connection with $\tau$-decay 
data~\cite{Barate:1998uf,Ackerstaff:1998yj} which give accurate information on 
the vector and axialvector spectral function in vacuum up to energies of the 
$\tau$ mass, $\sqrt{s} < m_\tau \simeq1.78$\,GeV. The low-energy part of the 
vector spectral function has been taken from the microscopic model for the $\rho$ 
meson that figures in the discussion of the dilepton data of the previous section, 
while the $a_1$ has been fit to the data with a Breit-Wigner ansatz. A novel 
element in this analysis is the postulate of a universal perturbative continuum 
in both vector and axialvector channel. Besides its underlying physical 
motivation of degeneracy in the perturbative domain, it also allows for an 
improved description of the intermediate-energy vector data through the 
introduction of a $\rho'$ resonance. It then turns out that a quantitative 
agreement with the right-hand-side ({\em rhs}) of WSRs 0-2 unequivocally 
requires the presence of an excited  axialvector resonance, at a mass of about 
$m_{a_1'}\simeq1.8$\,GeV (WSR-3 also shows a large improvement)~\cite{Hohler:2012xd}.   
This indicates a rather high sensitivity of the WSRs to chiral breaking effects. 
Another way of looking at this is to examine the values of the integrals on 
the {\it rhs} of the WSRs, Eq.~(\ref{wsr}), as a function of the upper integration 
limit, $I_n(s_{\rm up})$. One finds oscillations with an amplitude much larger 
than the asymptotic values as given by the left-hand side ({\it lhs}), i.e., the
latter are a result of formidable cancellations and therefore are to be considered 
as ``small".  Finally, the thus constructed vacuum spectral functions have been 
tested  with their respective QCDSR~\cite{Hohler:2012xd}. Within the current 
theoretical uncertainties of the additionally involved chiral blind operators, 
such as the gluon condensate, these are also reasonably well satisfied, within 
a typical margin of $\sim$0.5\%.

Let us now turn to the evaluation of the sum rules in medium (see, e.g., 
Ref.~\cite{Friman:2011zz} for a recent survey). As a first step, 
it is instructive to examine the consequences of model-independent low-temperature 
expansions.  As shown in Ref.~\cite{Dey:1990ba}, the leading $T$-dependence in 
the vector and axial-vector channels in the chiral limit amounts to a mutual 
mixing of their vacuum form, $\Pi_{V,A}^{0}$, as
\begin{equation}
\Pi_{V,A}(q;T) = (1-\varepsilon(T))~\Pi_{V,A}^{0}(q) + 
\varepsilon(T)~\Pi_{A,V}^{0}(q) \ , 
\label{chi-mix}
\end{equation}
with the mixing parameter $\varepsilon(T)= T^2/6f_\pi^2$ (which is proportional
to the scalar pion density). The physical process realizing the mixing is a 
resonant interaction of the $\rho$-meson with pions into an $a_1$, $\rho+\pi\to a_1$.
It turns out that the chiral mixing straightforwardly satisfies the
in-medium WSR-1 and -2, provided they are fulfilled in vacuum~\cite{Kapusta:1993hq}.
Complications can arise, however, at the level of the spectral functions, especially 
in set-ups with a rather small vector continuum threshold, where the latter does not 
separate
from the nonperturbative resonance region of the axialvector, i.e., the $a_1$ mass. 
This problem does not occur in the implementations of degenerate continua with 
higher threshold energy~\cite{Marco:2001dh,Hohler:2012xd}; here, the continua stay 
invariant and the mixing only operates on the nonperturbative part
of both spectral functions. One can furthermore evaluate the QCD 
sum rules in the mixing scenario~\cite{Holt:2012,Kwon:2010fw}). 
Employing the model-independent $T$-dependencies of quark and gluon 
condensates, one finds both vector and axialvector QCDSRs to be satisfied 
within 0.7\% or so up to temperatures of $T$=150\,MeV~\cite{Holt:2012}, where 
$\varepsilon\simeq0.2$ so that the expansion starts to become unreliable (note 
that $\varepsilon\simeq0.5$, which is reached at $T\simeq 225$\,MeV (160\,MeV 
in the chiral limit), corresponds to full mixing, i.e., degeneracy of $V$ and 
$A$ correlators in Eq.~(\ref{chi-mix}); thermal excitations other than the pion 
are expected to take over well before that). We recall that a mixing scenario 
can also be formulated in cold nuclear matter, through a coupling to the nuclear 
pion cloud~\cite{Krippa:1997ss,Chanfray:1999me}. The coupling of
these pions proceeds through the pion cloud of the $\rho$, thereby creating
axial currents~\cite{Chanfray:1999me}. 

\begin{figure}[!t]
\begin{minipage}{19pc}
\includegraphics[width=19pc]{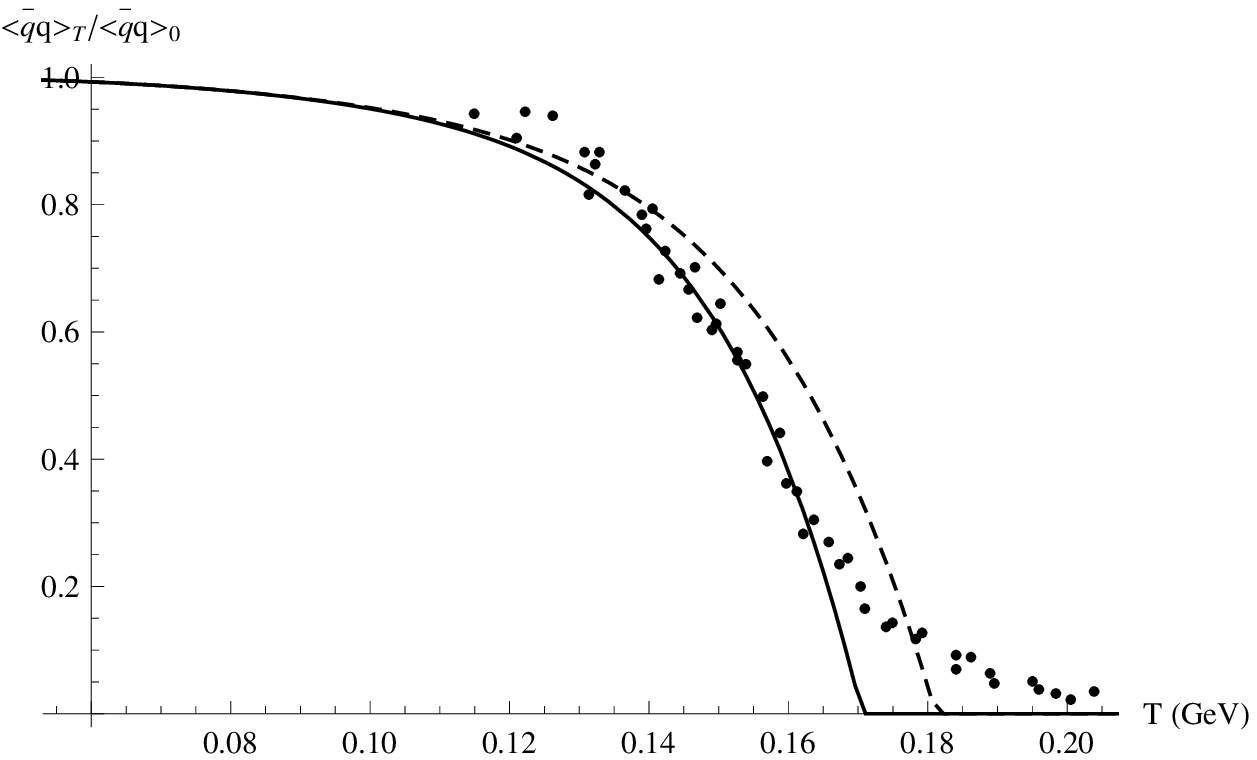}
\end{minipage}\hspace{1pc}%
\begin{minipage}{19pc}
\vspace{0.6pc}
\includegraphics[width=18pc]{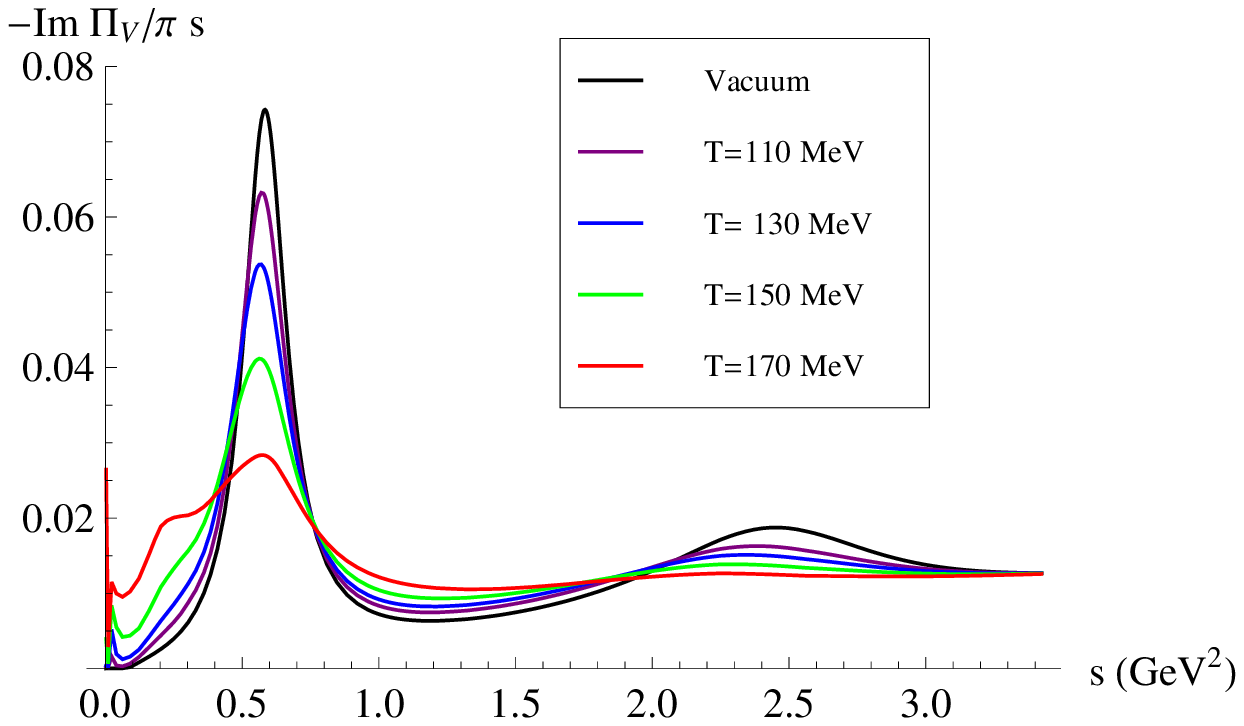}
\end{minipage}
\caption{Left panel: Temperature dependence of the chiral quark condensate, normalized
to its vacuum value, as obtained from thermal lattice QCD~\cite{Borsanyi:2010bp} for
different temporal lattice sizes (indicated by different symbols); the dashed line is 
the hadron-resonance gas result where the condensate is diminished by the quark content 
of the thermal excitations; the solid line includes a $T^{10}$ correction which
improves the agreement with lattice data at $T>$~140\,MeV.
Right panel: In-medium vector spectral functions in the isovector channel including 
the calculated in-medium $\rho$ contributions at vanishing chemical potential at low 
mass~\cite{Rapp:1999us}, supplemented with an in-medium $\rho'$ contribution and a 
fixed perturbative continuum as deduced from in-medium QCD sum rules~\cite{Hohler:2012fj}.}
\label{fig_med}
\end{figure}
While conceptually attractive (and rigorous), the $V$-$A$ mixing mechanism alone is 
insufficient to account for medium effects necessary to understand dilepton data. 
Most importantly, it lacks the broadening of the $\rho$ spectral shape that is 
pivotal in the description of the low-mass excess observed in experiment. We recall 
that this broadening is essentially due to two mechanisms~\cite{Rapp:1999us}: one is 
the medium modification of the $\rho$-meson's pion cloud (which includes virtual 
pion-exchange processes, i.e., chiral mixing), and the other is due to direct 
resonance interactions of the $\rho$ with heat bath particles $h$, $\rho+h \to R$, 
leading to the excitation of further resonances, $R$. How are these processes 
related to chiral symmetry restoration? To elaborate on this question, let us 
recall the model-independent, low-temperature and low-density result for the 
chiral condensate in a dilute hadronic medium (e.g., pion or nucleon gas) which 
reads~\cite{Gerber:1988tt,Drukarev:1991fs,Cohen:1991nk}
\begin{equation}
\frac{\langle\bar qq\rangle (T,\mu_B)}{\langle \bar qq\rangle_0} \ = \
1-\sum\limits_h \frac{\varrho_h^s \Sigma_h}{m_\pi^2 f_\pi^2}
\ \simeq \ 1 - \frac{T^2}{8f_\pi^2}  - \frac{1}{3}
\frac{\varrho_N}{\varrho_0} - \cdots  
\label{qqbar-med}
\end{equation}
with $\rho_h^s$: scalar density of hadron $h$. The sigma term,  
$\Sigma_h = m_q \langle h|\bar qq|h\rangle$, can be defined as the expectation value
of the chiral-symmetry breaking term of the QCD Lagrangian inside a hadron $h$. 
For the pion and nucleon they have been evaluated from both chiral perturbation 
theory~\cite{Gerber:1988tt,Drukarev:1991fs,Cohen:1991nk} and lattice 
QCD~\cite{MartinCamalich:2010fp}. The second equality in Eq.~(\ref{qqbar-med}) 
follows from the chiral limit for the pion and a value of $\Sigma_N$=45\,MeV for 
the nucleon.  More recent evaluations suggest the latter to be significantly larger, 
around 60\,MeV~\cite{MartinCamalich:2010fp}. The above expression has been 
generalized to a resonance gas of hadrons to leading order in their densities. 
Note that, although there are formally no interactions included, the excited 
hadronic states may be thought of as being built up by resonance interactions
of the stable pions and nucleons. In the nonrelativistic limit, the pertinent
sigma terms may be estimated using $\bar qq\simeq q^\dagger q$, so that 
$\Sigma_h/ m_q$ simply counts the number of light quarks in $h$. With $\Sigma_h>0$, 
one obtains the well-known result that the mere presence of hadrons diminishes the 
(negative) chiral condensate of the vacuum, 
$\langle 0|\bar qq|0\rangle\equiv\langle\bar qq\rangle_0\simeq-2$\,fm$^{-3}$ per 
light-quark flavor, i.e., the ``vacuum cleaner" effect.
The very existence of the resonance excitations in the hadron gas, which is one 
of the components in the $\rho$-meson broadening, is thus intimately related to 
the reduction of the chiral condensate. This notion can be carried further by 
realizing that the sigma term can be decomposed into a short-distance part 
associated with the hadron's quark core and a long-distance part associated 
with its pion cloud, $\Sigma_h = \Sigma_h^{\rm core} + \Sigma_h^{\pi}$, see, 
e.g., Refs.~\cite{Jameson:1992,Birse:1992} for the nucleon case. 
These two terms naturally find their counterparts in the medium effects of the
$\rho$ spectral function, namely the ones induced by direct resonance excitations 
as well as through the coupling of its pion cloud to the medium, respectively. 
This puts the medium effects due to chiral mixing through (virtual) pions and 
due to the hadron-resonance gas excitations on equal footing. Also recall
that the hadron-resonance gas appears to be a good approximation to the
QCD partition function for temperatures up to close to (and even slightly
above) $T_{\rm pc}$~\cite{Karsch:2003vd}.     

The next question is how this connection works out more quantitatively when analyzing 
the in-medium sum rules. This has been revisited very recently~\cite{Hohler:2012fj} 
(see also Ref.~\cite{Ayala:2012ch}), by adopting the newly 
suggested description of the vector correlator with ground and excited state and a 
universal high-energy continuum (recall Fig.~\ref{fig_vac}). For the ground state the 
in-medium $\rho$ spectral function as used in dilepton calculations has been employed, 
while the perturbative continuum remains unchanged. This leaves the Breit-Wigner 
parameters of the in-medium $\rho'$ to be adjusted (in lieu of the continuum threshold 
in previous QCDSR analyses). For the temperature dependence of the condensates, 
the ``non-interacting" hadron-resonance gas expression, Eq.~(\ref{qqbar-med}), has been 
used for the 2-quark condensate, which turns out to agree rather well with lattice data,
cf.~left panel of Fig.~\ref{fig_med} (a small correction has been introduced to better
reproduce the lattice data in the vicinity of $T_{\rm pc}^\chi$). 
The 4-quark and gluon condensate are treated
analogously, where the former includes correction terms to vanish at the same temperature 
as the 2-quark condensate. It turns out that the QCDSRs can be rather well satisfied even
until close to the vanishing of the quark condensates, provided that the $\rho'$ also 
melts, cf.~right panel of Fig.~\ref{fig_med}. Of course, one should keep in mind
that the vanishing of the quark condensates in the hadron-resonance gas is unrelated
to a real phase transition, although it may still indicate that this model captures
basic aspects of the medium when extrapolated close to $T_{\rm pc}$. This argument also 
applies to the in-medium calculations of the $\rho$ spectral function which should be 
rather reliable up to temperatures of about $T\simeq150$\,MeV, where the total hadron 
density has reached about 2$\varrho_0$. As discussed in the previous section, it is 
quite intriguing that dilepton data support a smooth extrapolation of the medium effects 
into the transition region. 
The analyses of the WSRs in this framework requires the knowledge of the in-medium
axialvector spectral function, which is not available (yet). Preliminary studies
using ans\"atze for the in-medium Breit-Wigner shape for $a_1$ and $a_1'$ that 
satisfy the axialvector QCDSR, indicate agreement with the in-medium WSRs, with 
a tendency of the vector and axialvector to degenerate into rather structureless 
spectral functions. 

\begin{figure}[!t]
\begin{minipage}{17pc}
\includegraphics[width=17pc]{Pii180.eps}
\vspace{-1.5pc}
\end{minipage}
\hspace{1pc}
\begin{minipage}{17pc}
\vspace{-1.5pc}
\includegraphics[width=16.3pc,angle=-90]{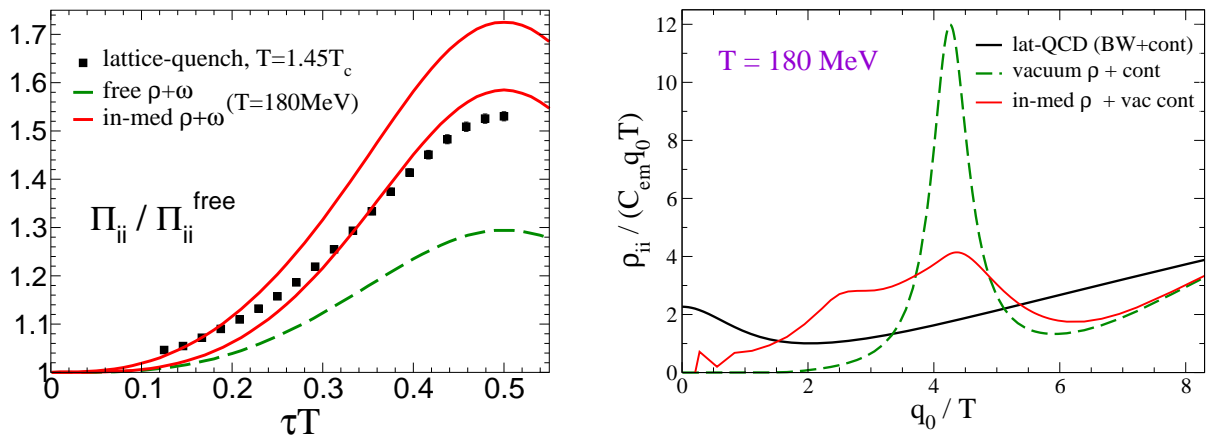}
\end{minipage}
%\vspace{0.1cm}
\caption{Left panel: Euclidean correlators at vanishing 
3-momentum and normalized to the free $q \bar q$ continuum (for $N_f$=2 light 
quarks), as a function of imaginary time, $\tau$, in units of inverse temperature;
the thermal lattice QCD results in quenched approximation at 1.45\,$T_c$ (black 
squares)~\cite{Ding:2010ga} are compared to effective hadronic model 
evaluations of Eq.~(\ref{Pi-tau2})~\cite{Rapp:2002pn} using either vacuum $\rho$
and $\omega$ spectral functions plus continuum (dashed line), or in-medium $\rho$
and $\omega$~\cite{Rapp:1999us} at $T$=180\,MeV with either vacuum or in-medium 
reduced continuum (lower and upper solid line, respectively).
Right panel: Spectral functions (normalized by isospin degeneracy, temperature and   
energy) corresponding to the correlators in the left panel; the lQCD result (black
solid line) is extracted from the data points in the right panel by a 3-parameter fit
ansatz~\cite{Ding:2010ga}; for the hadronic spectral functions only the isovector
part is shown ($\rho$ meson plus continuum), where the in-medium one (red line) uses 
a vacuum continuum and thus corresponds to the lower red line in the left panel.}
\label{fig_lat}
\end{figure}
Another test of the in-medium vector spectral function, and of the associated
chiral-restoration mechanism, can be provided by thermal
lattice QCD. In the latter, the information on the correlation function in a
specific hadronic channel, $\alpha$, is routinely computed in euclidean time, 
$\Pi_\alpha(\tau,\vec r)$. After a Fourier transform in the spatial coordinates
(from $\vec r$ to $\vec q$), the relation to the spectral function in the physical
(timelike) regime takes the form
\begin{equation}
\Pi_\alpha(\tau,q;T)=
\int\limits_0^\infty \frac{dq_0}{2\pi} \ \rho_\alpha(q_0,q;T) \
\frac{\cosh[q_0(\tau-1/2T)]}{\sinh[q_0/2T]} \ .
\label{Pi-tau2}
\end{equation} 
Thus, via simple a integration over a model spectral function one can directly
compare to ``lattice data". The computational effort to compute the euclidean
correlators in full QCD with dynamical quarks is formidable and no results for
light vector mesons are currently available. However, in the quenched 
approximation, these calculations have achieved good accuracy in a gluon plasma 
above $T_c$~\cite{Ding:2010ga}. It is instructive to compare the trends in these
computations to what one obtains from the spectral functions used in the 
interpretation of dilepton data. This is shown in the left panel of 
Fig.~\ref{fig_lat}, depicting the euclidean correlators normalized to the 
perturbative (non-interacting) $q\bar q$ continuum. The hadronic spectral function, 
extrapolated to a temperature of $T$=180\,MeV, shows a significant enhancement
at large $\tau$ which is a direct manifestation of the low-mass enhancement
generated by medium effects (critical for the description of dilepton data).
In fact, even the vacuum spectral function shows this effect, caused by the
free $\rho$ and $\omega$ resonances; the in-medium broadening roughly doubles
this enhancement. The quenched lattice data~\cite{Ding:2010ga} show a surprisingly 
similar trend, given that they are computed at 1.45\,$T_c$. Note that chiral symmetry
is restored under these conditions~\cite{Boyd:1995cw}. To extract the spectral
function from the lattice data is more involved. In Ref.~\cite{Ding:2010ga} a 
physically motivated 3-parameter ansatz has been fit to the correlators resulting
in a conductivity maximum at low-energy, followed by a smooth transition into
the perturbative continuum at high energy (see black solid line in the right
panel of Fig.~\ref{fig_lat}). Comparing this to the hadronic spectral functions, 
one observes a trend suggestive of approaching the lattice data via a 
melting of the vacuum $\rho$ resonance structure.

%%%%%%%%%%%%%%%%%%%%%%%%%%%%%%%%%%%%%%%%%%%%%
\section{Conclusions}
\label{sec_concl}
%%%%%%%%%%%%%%%%%%%%%%%%%%%%%%%%%%%%%%%%%%%%%
Low-mass dilepton data in ultrarelativistic heavy-ion collisions provide a unique 
glimpse at the vector spectral function inside the produced hot and dense medium.
After the initial discovery of large medium effects in the early and mid 90's,
recent data are now allowing for quantitative tests of theoretical calculations. 
At SPS energies, a strongly broadened $\rho$-meson spectral function, due to 
many-body effects in hot and dense hadronic matter, accounts well for both CERES 
and NA60 data.  Slope analyses of invariant-mass and transverse-momentum 
spectra corroborate that the observed spectral modifications at low mass originate 
from temperatures in the vicinity of the QCD transition region, 
$T_{\rm pc}$$\simeq$160-170\,MeV. If this picture is correct, very similar effects 
are expected for the low-mass dilepton spectra at collider energies. Very recent 
STAR data have given first evidence for the universality of the low-mass excess 
with collision energy, but are hopefully only the beginning of a systematic multi-differential 
investigation of dilepton observables. The transition from the baryon-rich to 
net-baryon free matter is a critical test of the current understanding of medium
effects driven by baryon plus anti-baryon densities in the vector spectral function. 
At the same time, the large PHENIX photon-$v_2$ is most naturally associated with 
radiation from around $T_{\rm pc}$, even at full RHIC energy. The large PHENIX 
dilepton excess in central Au-Au remains a puzzle which most likely requires a new 
type of radiation source. 

We have then argued that the mechanisms underlying these medium effects, namely 
resonance excitations and (virtual) pion-cloud modifications, find their direct 
counterpart in the sigma-terms of the heat-bath particles. This is significant as the
sigma-terms are at the origin of the reduction of the chiral quark condensate. Even
though this mechanism only captures the leading dependence in the scalar density
of the medium particles, it is remarkable that this ansatz describes the lattice
data for the quark condensate rather well until close to the (pseudo-) transition 
temperature. Although the resonance gas does not incorporate criticality, the 
resonance excitations represent interaction contributions which amount to higher 
orders in the density of the stable pions and nucleons.  
%Given that the hadron-resonance gas also appears to give a good
%description of QCD thermodynamics until close to $T_c$, one is led to conlcude that
%the $\rho$ melting is conistent 
When evaluating the relation between the in-medium vector spectral function and 
the decreasing condensate(s) more quantitatively using QCD sum rules, a reasonable
consistency is found; studies of the Weinberg sum rules are ongoing. Finally, 
the calculations of euclidean correlators and their comparison to current lattice
data also reveal a common trend toward a rather structureless spectral function.      
All this suggests that the $\rho$-meson melting scenario is quite consistent with
different angles on chiral symmetry restoration. To develop these indications into a 
quantitative proof, and/or unravel hitherto unknown aspects of chiral restoration, 
remains a challenge. Experimental information from the collider experiments will be 
critical in guiding further theoretical efforts, and vice versa.

\vspace{1pc}

\noindent {\bf Acknowledgment}
I thank H.~van Hees, C.~Gale and J.~Wambach for fruitful collaboration, and 
P.~Hohler and N.~Holt for their recent contributions to the sum rule analyses.
This work is supported by the US National Science Foundation under
grant no.~PHY-0969394 and by the A.-v.-Humboldt foundation.

%%%%%%%%%%%%%%%%%%%%%%%%%%%%%%%%%%%%%%%%


\section*{References}
\begin{thebibliography}{9}
%%%%%%%%%%%%%%%%%%%%%%%%%%%%%%%%%%%%%%%

\bibitem{Rapp:2009yu}
  R.~Rapp, J.~Wambach and H.~van Hees,
  %``The Chiral Restoration Transition of QCD and Low Mass Dileptons,''
  in {\em Relativistic Heavy-Ion Physics}, edited by R.~Stock and
  Landolt B\"ornstein (Springer), New Series {\bf I/23A} (2010) 4-1
  [arXiv:0901.3289[hep-ph]].
  %%CITATION = ARXIV:0901.3289;%%


\bibitem{Tserruya:2009zt}
  I.~Tserruya,
 %``Electromagnetic Probes,''
in {\em Relativistic Heavy-Ion Physics}, edited by R.~Stock and
  Landolt B\"ornstein (Springer), New Series {\bf I/23A} (2010) 4-2
  [arXiv:0903.0415[nucl-ex]].
  %%CITATION = ARXIV:0903.0415;%%

\bibitem{Specht:2010xu}
  H.J.~Specht  [for the NA60 Collaboration],
  %``Thermal Dileptons from Hot and Dense Strongly Interacting Matter,''
  AIP Conf.\ Proc.\  {\bf 1322}, 1 (2010).
%  [arXiv:1011.0615 [nucl-ex]].
  %%CITATION = APCPC,1322,1;%%

\bibitem{Rapp:2011is} 
  R.~Rapp,
  %``Theory of Soft Electromagnetic Emission in Heavy-Ion Collisions,''
  Acta Phys.\ Polon.\ B {\bf 42}, 2823 (2011).
%  [arXiv:1110.4345 [nucl-th]].
  %%CITATION = ARXIV:1110.4345;%%

\bibitem{Gale:2012xq} 
  C.~Gale,
  %``Electromagnetic radiation in heavy ion collisions: Progress and puzzles,''
  arXiv:1208.2289 [hep-ph].
  %%CITATION = ARXIV:1208.2289;%%


\bibitem{Agakishiev:1995xb} 
  G.~Agakichiev {\it et al.}  [CERES Collaboration],
%``Enhanced Production Of Low Mass Electron Pairs In 200-gev/u S - Au Collisions At The Cern Sps"
  Phys.\ Rev.\ Lett.\  {\bf 75}, 1272 (1995).
  %%CITATION = PRLTA,75,1272;%%

\bibitem{Agakichiev:2005ai} 
  G.~Agakichiev G {\it et al.}  [CERES Collaboration],
   %``e+ e- pair production in Pb - Au collisions at 158-GeV per nucleon,''
  {\it Eur.\ Phys.\ J.\ C} {\bf 41}, 475 (2005). 
%  [nucl-ex/0506002].
  %%CITATION = NUCL-EX/0506002;%%

\bibitem{Adamova:2006nu}
  D.~Adamova {\it et al.} [CERES/NA45 Collaboration],
  %``Modification of the rho meson detected by low-mass electron-positron pairs
  %in central Pb-Au collisions at 158 A GeV/c,''
  Phys.\ Lett.\  B {\bf 666}, 425 (2008).
%  [arXiv:nucl-ex/0611022].
  %%CITATION = PHLTA,B666,425;%%

\bibitem{Li:1995qm} 
  G.-Q.~Li, C.M.~Ko and G.E.~Brown,
  %``Enhancement of low mass dileptons in heavy ion collisions,''
  Phys.\ Rev.\ Lett.\  {\bf 75}, 4007 (1995).
%  [nucl-th/9504025].
  %%CITATION = NUCL-TH/9504025;%%

\bibitem{Brown:2009az} 
  G.E.~Brown, M.~Harada, J.W.~Holt, M.~Rho and C.~Sasaki,
  %``Hidden Local Field Theory and Dileptons in Relativistic Heavy Ion Collisions,''
  Prog.\ Theor.\ Phys.\  {\bf 121}, 1209 (2009)
%  [arXiv:0901.1513 [hep-ph]].
  %%CITATION = ARXIV:0901.1513;%%

\bibitem{Chanfray:1993ue} 
  G.~Chanfray and P.~Schuck,
  %``The Rho meson in dense matter and its influence on dilepton production rates,''
  Nucl.\ Phys.\ A {\bf 555}, 329 (1993).
  %%CITATION = NUPHA,A555,329;%%

\bibitem{Herrmann:1993za} 
  M.~Herrmann, B.L.~Friman and W.~N\"orenberg,
  %``Properties of rho mesons in nuclear matter,''
  Nucl.\ Phys.\ A {\bf 560}, 411 (1993).
  %%CITATION = NUPHA,A560,411;%%

\bibitem{Friman:1997tc} 
  B.~Friman and H.J.~Pirner,
  %``P wave polarization of the rho meson and the dilepton spectrum in dense matter,''
  Nucl.\ Phys.\ A {\bf 617}, 496 (1997).
%  [nucl-th/9701016].
  %%CITATION = NUCL-TH/9701016;%%

\bibitem{Klingl:1997kf} 
  F.~Klingl, N.~Kaiser and W.~Weise,
  %``Current correlation functions, QCD sum rules and vector mesons in baryonic matter,''
  Nucl.\ Phys.\ A {\bf 624}, 527 (1997).
%  [hep-ph/9704398].
  %%CITATION = HEP-PH/9704398;%%

\bibitem{Peters:1997va} 
  W.~Peters, M.~Post, H.~Lenske, S.~Leupold and U.~Mosel,
  %``The Spectral function of the rho meson in nuclear matter,''
  Nucl.\ Phys.\ A {\bf 632}, 109 (1998).
% [nucl-th/9708004].
  %%CITATION = NUCL-TH/9708004;%%

\bibitem{Urban:1998eg} 
  M.~Urban, M.~Buballa, R.~Rapp and J.~Wambach,
  %``Momentum dependence of the pion cloud for rho mesons in nuclear matter,''
  Nucl.\ Phys.\ A {\bf 641}, 433 (1998).
%  [nucl-th/9806030].
  %%CITATION = NUCL-TH/9806030;%%

\bibitem{Cabrera:2000dx} 
  D.~Cabrera, E.~Oset and M.~J.~Vicente Vacas,
  %``Chiral approach to the rho meson in nuclear matter,''
  Nucl.\ Phys.\ A {\bf 705}, 90 (2002).
%  [nucl-th/0011037].
  %%CITATION = NUCL-TH/0011037;%%


\bibitem{Haglin:1994xu} 
  K.~Haglin,
  %``Collision rates for rho, omega and phi mesons at nonzero temperature,''
  Nucl.\ Phys.\ A {\bf 584}, 719 (1995).
%  [nucl-th/9410028].
  %%CITATION = NUCL-TH/9410028;%%

\bibitem{Pisarski:1995xu} 
  R.~D.~Pisarski,
  %``Where does the rho go? Chirally symmetric vector mesons in the quark - gluon plasma,''
  Phys.\ Rev.\ D {\bf 52}, 3773 (1995).
%  [hep-ph/9503328].
  %%CITATION = HEP-PH/9503328;%%

\bibitem{Song:1996dg} 
  C.~Song and V.~Koch,
  %``Pion electromagnetic form-factor at finite temperature,''
  Phys.\ Rev.\ C {\bf 54}, 3218 (1996).
%  [nucl-th/9608010].
  %%CITATION = NUCL-TH/9608010;%%

\bibitem{Rapp:1999qu} 
  R.~Rapp and C.~Gale,
  %``Rho properties in a hot gas: Dynamics of meson resonances,''
  Phys.\ Rev.\ C {\bf 60}, 024903 (1999).
%  [hep-ph/9902268].
  %%CITATION = HEP-PH/9902268;%%

\bibitem{Ayala:2003yp} 
  A.~Ayala and J.~Magnin,
  %``Rho propagation and dilepton production at finite pion density and temperature,''
  Phys.\ Rev.\ C {\bf 68}, 014902 (2003)
%  [hep-ph/0302152].
  %%CITATION = HEP-PH/0302152;%%


\bibitem{Rapp:1995zy} 
  R.~Rapp, G.~Chanfray and J.~Wambach,
  %``Medium modifications of the rho meson at CERN SPS energies,''
  Phys.\ Rev.\ Lett.\  {\bf 76}, 368 (1996).
%  [hep-ph/9508353].
  %%CITATION = HEP-PH/9508353;%%

\bibitem{Rapp:1999us}
R.~Rapp and J.~Wambach, Eur. Phys. J. A {\bf 6}, 415 (1999).
%[{\tt hep-ph/9907502}].
%%CITATION = EPHJA,A6,415;%%

\bibitem{Eletsky:2001bb} 
  V.L.~Eletsky, M.~Belkacem, P.J.~Ellis and J.I.~Kapusta,
%``Properties of rho and omega mesons at finite temperature and density as inferred from experiment,''
  Phys.\ Rev.\ C {\bf 64}, 035202 (2001).
%  [nucl-th/0104029].
  %%CITATION = NUCL-TH/0104029;%%

\bibitem{Ghosh:2011gs}
  S.~Ghosh and S.~Sarkar,
  %``$\rho$ self energy at finite temperature and density in the real-time formalism,''
  Nucl.\ Phys.\ A {\bf 870-871}, 94 (2011)
  [Erratum-ibid.\ A {\bf 888}, 44 (2012)].
%  [arXiv:1109.2773 [nucl-th]].
  %%CITATION = ARXIV:1109.2773;%%

\bibitem{Arnaldi:2008fw}
  R.~Arnaldi R {\it et al.}  [NA60 Collaboration],
%``NA60 results on thermal dimuons,''
 {\it Eur. Phys. J. C} {\bf 61}, 711 (2009).
%[{\tt 0812.3053[nucl-ex]}].

\bibitem{vanHees:2007th} 
  H.~van Hees and R.~Rapp,
  %``Dilepton Radiation at the CERN Super Proton Synchrotron,''
  Nucl.\ Phys.\ A {\bf 806}, 339 (2008)
%  [arXiv:0711.3444 [hep-ph]].
  %%CITATION = ARXIV:0711.3444;%%

\bibitem{Arnaldi:2006jq}
R.~Arnaldi et~al. [NA60 Collaboration], Phys. Rev. Lett. \textbf{96}, 162302 (2006).

\bibitem{vanHees:2006ng} 
  H.~van Hees and R.~Rapp,
  %``Comprehensive interpretation of thermal dileptons at the SPS,''
  Phys.\ Rev.\ Lett.\  {\bf 97}, 102301 (2006)
  [hep-ph/0603084].
  %%CITATION = HEP-PH/0603084;%%

\bibitem{Dusling:2006yv} 
  K.~Dusling, D.~Teaney and I.~Zahed,
  %``Thermal dimuon yields at NA60,''
  Phys.\ Rev.\ C {\bf 75}, 024908 (2007)
  [nucl-th/0604071].
  %%CITATION = NUCL-TH/0604071;%%

\bibitem{Ruppert:2007cr} 
  J.~Ruppert, C.~Gale, T.~Renk, P.~Lichard and J.~I.~Kapusta,
  %``Low mass dimuons produced in relativistic nuclear collisions,''
  Phys.\ Rev.\ Lett.\  {\bf 100}, 162301 (2008)
%  [arXiv:0706.1934 [hep-ph]].
  %%CITATION = ARXIV:0706.1934;%%

\bibitem{Santini:2011zw} 
  E.~Santini, J.~Steinheimer, M.~Bleicher and S.~Schramm,
  %``Dimuon radiation at the CERN SPS within a (3+1)d hydrodynamic+cascade model,''
  Phys.\ Rev.\ C {\bf 84}, 014901 (2011)
%  [arXiv:1102.4574 [nucl-th]].
  %%CITATION = ARXIV:1102.4574;%%

\bibitem{Linnyk:2011hz} 
  O.~Linnyk, E.L.~Bratkovskaya, V.~Ozvenchuk, W.~Cassing and C.M.~Ko,
 %``Dilepton production in nucleus-nucleus collisions at top SPS energy within 
 %the Parton-Hadron-String Dynamics (PHSD) transport approach,''
  Phys.\ Rev.\ C {\bf 84}, 054917 (2011)
%  [arXiv:1107.3402 [nucl-th]].
  %%CITATION = ARXIV:1107.3402;%%

\bibitem{Rapp:1999if}
  R.~Rapp,
  %``Duality and chiral restoration from low mass dileptons at the CERN SPS,''
  Nucl.\ Phys.\ A {\bf 661}, 33 (1999).
%  [hep-ph/9907342].
  %%CITATION = HEP-PH/9907342;%%

\bibitem{Kapusta:1991qp} 
  J.I.~Kapusta, P.~Lichard and D.~Seibert,
  %``High-energy photons from quark - gluon plasma versus hot hadronic gas,''
  Phys.\ Rev.\ D {\bf 44}, 2774 (1991)
  [Erratum-ibid.\ D {\bf 47}, 4171 (1993)].
  %%CITATION = PHRVA,D44,2774;%%

\bibitem{Turbide:2003si}
  S.~Turbide, R.~Rapp and C.~Gale,
  %``Hadronic production of thermal photons,''
  Phys.\ Rev.\  C {\bf 69}, 014903 (2004).
%  [arXiv:hep-ph/0308085].
  %%CITATION = PHRVA,C69,014903;%%

\bibitem{Arnaldi:2007ru}
  R.~Arnaldi {\it et al.}  [NA60 Collaboration],
  %``Evidence for radial flow of thermal dileptons in high-energy nuclear collisions,''
  Phys.\ Rev.\ Lett.\  {\bf 100}, 022302 (2008).
%  [arXiv:0711.1816 [nucl-ex]].
  %%CITATION = ARXIV:0711.1816;%%

\bibitem{Rapp:1997ei}
R.~Rapp, M.~Urban, M.~Buballa, and J.~Wambach,
Phys. Lett. B \textbf{417}, 1 (1998).

%\bibitem{Braaten:1990wp}
%E.~Braaten, R.D.~Pisarski, and T.-C.~Yuan, Phys. Rev. Lett. \textbf{64}, 2242 (1990).

\bibitem{Adamova:2003kf}
D.~Adamova et~al. [CERES/NA45 Collaboration], Phys. Rev. Lett. \textbf{91},
042301 (2003).

\bibitem{Porter:1997rc}
R.J.~Porter et~al. (DLS Collaboration), Phys. Rev. Lett. \textbf{79}, 1229 (1997).

\bibitem{Agakishiev:2011vf} 
  G.~Agakishiev {\it et al.}  [HADES Collaboration],
  %``Dielectron production in Ar+KCl collisions at 1.76A GeV,''
  Phys.\ Rev.\ C {\bf 84}, 014902 (2011)
%  [arXiv:1103.0876 [nucl-ex]].
  %%CITATION = ARXIV:1103.0876;%%

\bibitem{Rapp:2000pe} 
  R.~Rapp,
  %``Signatures of thermal dilepton radiation at RHIC,''
  Phys.\ Rev.\ C {\bf 63}, 054907 (2001).
%  [hep-ph/0010101].
  %%CITATION = HEP-PH/0010101;%%

\bibitem{Adare:2009qk}
  A.~Adare {\it et al.}  [PHENIX Collaboration],
  %``Detailed measurement of the $e^+ e^-$ pair continuum in $p+p$ and Au+Au
  %collisions at $\sqrt{s_{NN}} = 200$ GeV and implications for direct photon
  %production,''
  Phys.\ Rev.\  C {\bf 81}, 034911 (2010).
%  [arXiv:0912.0244 [nucl-ex]].
  %%CITATION = PHRVA,C81,034911;%%

\bibitem{Zhao:2011wa} 
  J.~Zhao [STAR Collaboration],
  %``Dielectron continuum production from $\sqrt{s_{NN}}$ = 200 GeV p + p and Au + Au collisions at STAR,''
  J.\ Phys.\ G {\bf 38}, 124134 (2011).
%  [arXiv:1106.6146 [nucl-ex]].
  %%CITATION = ARXIV:1106.6146;%%

\bibitem{Geurts:2012}
F.~Geurts {\it et al.} [STAR Collaboration], talk at Quark Mater 2012 International Conference
(Washington, DC), Aug. 13-18, 2012.

\bibitem{Tserruya:2012}
I.~Tserruya {\it et al.} [PHENIX Collaboration], talk at Quark Mater 2012 International Conference
(Washington, DC), Aug. 13-18, 2012.

\bibitem{Adare:2011zr} 
  A.~Adare {\it et al.}  [PHENIX Collaboration],
  %``Observation of direct-photon collective flow in sqrt(s_NN)=200 GeV Au+Au collisions,''
  arXiv:1105.4126 [nucl-ex].
  %%CITATION = ARXIV:1105.4126;%%

\bibitem{Liu:2009kta}
  F.~-M.~Liu, T.~Hirano, K.~Werner and Y.~Zhu,
  %``Elliptic flow of thermal photons in Au + Au collisions at s(NN)**(1/2) = 200-GeV,''
  Phys.\ Rev.\ C {\bf 80}, 034905 (2009).
%  [arXiv:0902.1303 [hep-ph]].
  %%CITATION = ARXIV:0902.1303;%%

\bibitem{Holopainen:2011pd}
  H.~Holopainen, S.~Rasanen and K.~J.~Eskola,
%``Elliptic flow of thermal photons in heavy-ion collisions at Relativistic Heavy Ion Collider and Large Hadron
Collider,''
  Phys.\ Rev.\ C {\bf 84}, 064903 (2011).
%  [arXiv:1104.5371 [hep-ph]].
  %%CITATION = ARXIV:1104.5371;%%

\bibitem{Dion:2011pp}
  M.~Dion, J.~-F.~Paquet, B.~Schenke, C.~Young, S.~Jeon and C.~Gale,
  %``Viscous photons in relativistic heavy ion collisions,''
  Phys.\ Rev.\ C {\bf 84}, 064901 (2011).
%  [arXiv:1109.4405 [hep-ph]].
  %%CITATION = ARXIV:1109.4405;%%

\bibitem{vanHees:2011vb} 
  H.~van Hees, C.~Gale and R.~Rapp,
  %``Thermal Photons and Collective Flow at the Relativistic Heavy-Ion Collider,''
  Phys.\ Rev.\ C {\bf 84}, 054906 (2011).
%  [arXiv:1108.2131 [hep-ph]].
  %%CITATION = ARXIV:1108.2131;%%

\bibitem{Arnold:2001ms}
P.B.~Arnold, G.D.~Moore and L.G.~Yaffe,
%``Photon emission from quark gluon plasma: Complete leading order results,''
  JHEP {\bf 0112}, 009 (2001).
%  [arXiv:hep-ph/0111107].
  %%CITATION = JHEPA,0112,009;%%

\bibitem{He:2011zx} 
  M.~He, R.J.~Fries and R.~Rapp,
  %``Ideal Hydrodynamics for Bulk and Multistrange Hadrons in $\sqrt{s_{NN}}$=200\,AGeV Au-Au Collisions,''
  Phys.\ Rev.\ C {\bf 85}, 044911 (2012).
%  [arXiv:1112.5894 [nucl-th]].
  %%CITATION = ARXIV:1112.5894;%%

%\bibitem{Basar:2012bp} 
%  G.~Basar, D.~Kharzeev and V.~Skokov,
  %``Conformal anomaly as a source of soft photons in heavy ion collisions,''
%  arXiv:1206.1334 [hep-ph].
  %%CITATION = ARXIV:1206.1334;%%

\bibitem{Shifman:1978bx}
M.A.~Shifman, A.I.~Vainshtein, and V.I.~Zakharov, Nucl. Phys. B
  \textbf{147}, 385 (1979).

\bibitem{Weinberg:1967}
S.~Weinberg, Phys. Rev. Lett. \textbf{18}, 507 (1967).
                                                                                
\bibitem{Das:1967ek}
T.~Das, V.~S. Mathur, and S.~Okubo, Phys. Rev. Lett. \textbf{19}, 859 (1967).

\bibitem{Kapusta:1993hq} 
  J.I.~Kapusta and E.V.~Shuryak,
  %``Weinberg type sum rules at zero and finite temperature,''
  Phys.\ Rev.\ D {\bf 49}, 4694 (1994).
%  [hep-ph/9312245].
  %%CITATION = HEP-PH/9312245;%%

\bibitem{Barate:1998uf}
R.~Barate et~al. (ALEPH Collaboration), Eur. Phys. J. C \textbf{4}, 409 (1998).
                                                                                
\bibitem{Ackerstaff:1998yj}
K.~Ackerstaff et~al. (OPAL Collaboration), Eur. Phys. J. C \textbf{7}, 571
  (1999).

\bibitem{Hohler:2012xd} 
  P.M.~Hohler and R.~Rapp,
%``Sum rule analysis of vector and axial-vector spectral functions with 
% excited states in vacuum,''
  Nucl. Phys. A {\bf 892},  58 (2012).
  %%CITATION = ARXIV:1204.6309;%%

\bibitem{Friman:2011zz} 
 B.~Friman, (ed.), C.~Hohne, (ed.), J.~Knoll, (ed.), S.~Leupold, (ed.), J.~Randrup, (ed.), 
R.~Rapp, (ed.) and P.~Senger, (ed.),
  %``The CBM physics book: Compressed baryonic matter in laboratory experiments,''
  Lect.\ Notes Phys.\  {\bf 814}, 1 (2011).
  %%CITATION = LNPHA,814,1;%%

\bibitem{Dey:1990ba}
M.~Dey, V.~L. Eletsky, and B.~L. Ioffe, Phys. Lett. B \textbf{252}, 620 (1990).

%\bibitem{Toublan:1997rr} 
%  D.~Toublan,
  %``Pion dynamics at finite temperature,''
%  Phys.\ Rev.\ D {\bf 56}, 5629 (1997)
%  [hep-ph/9706273].
  %%CITATION = HEP-PH/9706273;%%

\bibitem{Marco:2001dh} 
  E.~Marco, R.~Hofmann and W.~Weise,
  %``Note on finite temperature sum rules for vector and axial vector spectral functions,''
  Phys.\ Lett.\ B {\bf 530}, 88 (2002).
%  [hep-ph/0110110].
  %%CITATION = HEP-PH/0110110;%%

\bibitem{Holt:2012}
N.~Holt, P.M.~Hohler and R.~Rapp, in preparation (2012).

\bibitem{Kwon:2010fw}
  Y.~Kwon, C.~Sasaki and W.~Weise,
  %``Vector mesons at finite temperature and QCD sum rules,''
  Phys.\ Rev.\ C {\bf 81}, 065203 (2010).
 % [arXiv:1004.1059 [nucl-th]].
  %%CITATION = ARXIV:1004.1059;%%


\bibitem{Krippa:1997ss}
B.~Krippa, Phys. Lett. B \textbf{427}, 13 (1998).

\bibitem{Chanfray:1999me} 
  G.~Chanfray, J.~Delorme, M.~Ericson and M.~Rosa-Clot,
  %``Pion number and correlator mixing,''
  Phys.\ Lett.\ B {\bf 455}, 39 (1999).
  %%CITATION = PHLTA,B455,39;%%

\bibitem{Gerber:1988tt}
  P.~Gerber and H.~Leutwyler,
  %``Hadrons Below the Chiral Phase Transition,''
  Nucl.\ Phys.\ B {\bf 321}, 387 (1989).
  %%CITATION = NUPHA,B321,387;%%

\bibitem{Drukarev:1991fs}
E.~G. Drukarev and E.~M. Levin, Prog. Part. Nucl. Phys. \textbf{27}, 77 (1991).

\bibitem{Cohen:1991nk}
T.~D. Cohen, R.~J. Furnstahl, and D.~K. Griegel, Phys. Rev. C \textbf{45}, 1881
  (1992).

\bibitem{MartinCamalich:2010fp} 
  J.~Martin Camalich, L.~S.~Geng and M.~J.~Vicente Vacas,
  %``The lowest-lying baryon masses in covariant SU(3)-flavor chiral perturbation theory,''
  Phys.\ Rev.\ D {\bf 82}, 074504 (2010).
%  [arXiv:1003.1929 [hep-lat]].
  %%CITATION = ARXIV:1003.1929;%%

\bibitem{Jameson:1992}
I.~Jameson, A.W.~Thomas and G.~Chanfray, J. Phys. G {\bf 18}, L159 (1992).

\bibitem{Birse:1992}
M.C.~Birse and J.A.~McGovern, Phys. Lett. {\bf B292}, 242 (1992).

\bibitem{Karsch:2003vd} 
  F.~Karsch, K.~Redlich and A.~Tawfik,
  %``Hadron resonance mass spectrum and lattice QCD thermodynamics,''
  Eur.\ Phys.\ J.\ C {\bf 29}, 549 (2003).
%  [hep-ph/0303108].
  %%CITATION = HEP-PH/0303108;%%

\bibitem{Hohler:2012fj} 
  P.M.~Hohler and R.~Rapp,
  %``Evaluating chiral symmetry restoration through the use of sum rules,''
  arXiv:1209.1051 [hep-ph].
  %%CITATION = ARXIV:1209.1051;%%

\bibitem{Ayala:2012ch}
  A.~Ayala, C.~A.~Dominguez, M.~Loewe and Y.~Zhang,
  %``Rho-meson resonance broadening in QCD at finite temperature,''
  arXiv:1210.2588 [hep-ph].
  %%CITATION = ARXIV:1210.2588;%%

\bibitem{Borsanyi:2010bp}
  S.~Borsanyi, Z.~Fodor, C.~Hoelbling, S.D.~Katz, S.~Krieg, C.~Ratti and K.K.~Szabo
 [Wuppertal-Budapest Collaboration],
%``Is there still any T_c mystery in lattice QCD? Results with physical masses
%in the continuum limit III,''
  JHEP {\bf 1009}, 073 (2010).
%  [arXiv:1005.3508 [hep-lat]].
  %%CITATION = JHEPA,1009,073;%%

\bibitem{Ding:2010ga}
  H.T.~Ding, A.~Francis, O.~Kaczmarek, F.~Karsch, E.~Laermann and W.~Soeldner,
  %``Thermal dilepton rate and electrical conductivity: An analysis of vector
  %current correlation functions in quenched lattice QCD,''
  Phys.\ Rev.\  D {\bf 83}, 034504 (2011).
%  [arXiv:1012.4963 [hep-lat]].
  %%CITATION = PHRVA,D83,034504;%%

\bibitem{Rapp:2002pn} 
  R.~Rapp,
  %``Hadrons in hot and dense matter,''
  Eur. Phys. J. A {\bf 18}, 459 (2003).
%  nucl-th/0209081.
  %%CITATION = NUCL-TH/0209081;%%


\bibitem{Boyd:1995cw} 
  G.~Boyd, S.~Gupta, F.~Karsch, E.~Laermann, B.~Petersson and K.~Redlich,
  %``Hadron properties just before deconfinement,''
  Phys.\ Lett.\ B {\bf 349}, 170 (1995). 
%  [hep-lat/9501029].
  %%CITATION = HEP-LAT/9501029;%%




%\bibitem{Birse:1993he} 
%  M.C.~Birse and J.A.~McGovern,
%  %``Does the nuclear medium react against chiral symmetry restoration?,''
%  Phys.\ Lett.\ B {\bf 309}, 231 (1993).
%  %%CITATION = PHLTA,B309,231;%%


\end{thebibliography}
\end{document}